\documentstyle[preprint,aps]{revtex}
\font \afont = cmmi10 scaled \magstep 2
\mathchardef \A="0861 \rm
\mathchardef \B="0862 \rm
\begin{document}
\draft
\preprint{QG-3.3-12/17/95}
\title{Quantum Geodesics}
\author{Daniel C. Galehouse}
\address{Physics Department, University of Akron, Akron, Ohio 44325}
\date{\today}
\maketitle
\begin{abstract}
Classical methods of differential geometry are used
to construct equations of motion for particles
in quantum, electrodynamic and gravitational fields.
For a five dimensional geometrical system, the equivalence principle
can be extended.   Local transformations generate the effects of
electromagnetic and quantum fields.    A combination of
five dimensional coordinate transformations and
internal conformal transformations leads to a quantum Kaluza-Klein metric.
The theories of Weyl and Kaluza can be interrelated when charged particle
quantum mechanics is included.
Measurements of trajectories are made relative to an observers'
space that is defined by the motion of neutral particles.
It is shown that a preferred set of null geodesics
describe valid classical and quantum trajectories.  These are tangent to
the probability density four vector. This construction establishes a
generally covariant basis for geodesic motion of quantum states.
\end{abstract}
\pacs{MS: , PACS 03, 04.50 }
\vfill \eject
\narrowtext
\section{INTRODUCTION}
\label{int}

The fifth dimension has been discussed in the scientific literature
for seventy years or more.  Historically, the preponderance of attempts
to use five dimensions~\cite{r1,r2,r3,r3a,r4,r5,r6,r7,r8,r9,r9a,r10}
combine a gravitational theory similar to Einstein's~\cite{r11}
with an electromagnetic theory similar to Maxwell's~\cite{r12}.
The earliest versions were modeled after the general theory
of relativity~\cite{r13}, while later attempts to combine other
interactions~\cite{r14,r15,r16,r17}
are considerablely more intricate. The integration of quantum ideas
into theories has been frustrating and inconclusive.
Quantum mechanics is usually appended by  attaching a Hilbert space
to each region of space-time.

It is suggested here that there is a wholly geometrical
a way to describe physical phenomena, including quantum mechanics.
The historical exclusion of quantum mechanics from  fundamental classical
theories  is deemed arbitrary. It is proposed that the essential elements
of quantum mechanics reside within natural geometrical
structures~\cite{r18,r19,r1}.   Quantum mechanics becomes  an inherent,
inseparable part of the mathematics.

The basis of classical unified field theories rests on the work of
 H. Weyl \cite{r20} and  T. Kaluza \cite{r8}.  The  association
of these ideas with quantum mechanics has been know for some time.
Following earlier studies for the Weyl theories~\cite{r21,r19},
it has become possible to develop a five dimensional geometrical
quantum theory.

\section{INITIAL ASSUMPTIONS AND CONCEPTS}
\label{asu}

Several particular assumptions need to be discussed.  By stating unusual
starting points, it is hoped to save the reader the difficulty of inferring
them from the conclusions.

Much of the Einsteinian viewpoint is adopted.  Microscopic trajectories
are hypothesized as a universal description of the effects of
quantum mechanics, electromagnetism and general relativity.
While the success of this approach for quantum theory is not yet established,
it allows the direct discussion of gravitational phenomena.
In accord with the goal of a wholly geometrical theory, the
mathematics is hard deterministic and time symmetric.  Quantum probabilities
must be associated with a congruence of statistically populated
trajectories. Because of the metaphysical conflicts between quantum theory
and relativity, certain concepts  must be refashioned  to allow the synthesis.
The vector potential plays a more prominent role.   Metrics in
higher dimensions have modified interpretations.  The hilbert space
becomes a calculational artifact and not a starting point.

Electrodynamics is assumed time symmetric, along the lines of the
classical articles by Feynman and Wheeler~\cite{r22}.
For geodetic motion that includes electromagnetic effects,
the force of radiative reaction must be due
to fields derived from other particles.  There can be no force of
radiation until the interacting (absorbing) particles are included in the
field sources.  For a quantum gravitational theory, this time
symmetric construction must be supposed to exist at a
metaphysical level that is deeper than the macroscopic structure
of Maxwell.  Electrodynamics is assumed to derive from a primitive classical
truth of geometry, comparable with general relativity.

In a microscopically deterministic theory, the question of the separation
of causal agents from their effects
becomes problematical.  This does not appear to contradict the experimental
situation.   A precise
definition of causality probably cannot have a microscopic formulation
in a theory
that is hard deterministic, time symmetric and quantum mechanical.  Without
a primitive assumption, it is supposed that
time asymmetry and the associated causality follow from macroscopic
effects that may depend on other conditions in the universe.  These conditions
may include statistical mechanics,
cosmology, psychology or the real physical effects of other
interactions~\cite{r23}.
Microscopic time asymmetry is not eliminated in principle
but does not contribute to the discussion.

The geometrical concept of a congruence is associated with the
quantum mechanical state function of a particle.
Wave particle duality is not considered essential.
Since emission and absorption
is the only known evidence for photons, they
are assumed to be calculational artifacts that derive from
charged particle quantum mechanics~\cite{r24}.
Finite mass particles, having time like trajectories,
are the only true particles.
Interactions are mediated by fields. Any discreteness in the
fields (photons or gravitons) derives from the discreteness of the source
 particles~\cite{r25}.
It is assumed that the fields have no dynamical qualities except that
which is implied by the sources. Fundamental free fields are rejected.

To describe  more that one particle, the mathematical
field structure must be augmented.
Gravity  is conventionally approximated by a single
metric tensor that specifies  the global distortion of space-time.
In such an approximation, a collection of
particles $P_1,P_2, \dots P_k$ moves in a space-time specified
by a metric that is a solution of a set of gravitational field
equations.
This single metric is insufficient when constructing a combined
theory of electromagnetism and quantum mechanics.  The structures are
not analogous.  These latter effects are described by fields $A_{(1)\mu},
A_{(2)\mu},\dots A_{(k)\mu}$ and $\psi_{(1)},\psi_{(2)} \dots \psi_{(k)}$,
one each for each particle.   Of course the approximation of a universal
metric for a collection of simple particles is usually valid, particularly
in the laboratory.  However when quantum mechanics and electrodynamics are
integrated with gravity,
an analogous collection of metrics $g_{(1)\mu\nu},g_{(2)\mu\nu} \dots
g_{(k)\mu\nu}$ is required.
These metrics can be treated as equal under many experimental
situations; but in principle, even microscopic gravitational interactions
are present and provide interparticle forces. Consequently each particle
must have a formal wave function, vector potential and
metrical tensor, $(g_{\mu\nu},A_\mu,\psi)$.
An extended principle
of equivalence is made possible since the motion of each particle
can be the result of different combinations of assigned fields.
A special metric $\dot g_{\mu\nu}$, identified
with the observer of general relativity, is maintained separately from
the individual metrics of the particles.
It represents the structure of space-time as measured with neutral
particles and the idealized null trajectories of electromagnetic interactions.
Multiple metrics have been used at least since Dirac~\cite{r26}
and are implied by the conformal invariance of
Weyl~\cite{r20}.  Other multiple metric theories
of gravitation are listed by Ni~\cite{r27}.

A classical observer does not perceive
quantum mechanics or electromagnetic fields to be part of space-time
geometry. If these effects are to be
intrinsic, a separate phenomenology of charged and neutral particles must be
defined.  Hypothetical mirrors are assumed to be neutral objects to the
observer.  Forces of radiative reaction on a mirror are
neglected or compensated during gravitational space-time
measurements and the individual
charged particles which make up the mirror are treated collectively.
The observer's metric $\dot g_{\mu\nu}$ is used to describe the perceived
pattern of the collective motion of many particles.  As a mathematical
object, it  retains multiparticle
phenomenology and cannot be measured by using a single particle
wave function or quantum state.
Observations depend on idealized measurements of composite
particles and clocks that traverse each point with
 different velocities~\cite{r28}.
Each such independent direction  requires at least one
wave function~\cite{r29}.
Four dimensional space-time, must be derived in this way
from a larger geometry as a limit or approximation.  Stable, massive, neutral,
primitive (non-composite)
particles are required.  Since there are none known, the phenomenological
metric may
not have a basis in the
motion of elementary particles but may only have a well defined
meaning as a description  of collective motion.

An increase in dimensionality always leads to new
quantities and interpretations.  It is supposed that each
particle, no matter how it might be described ultimately,
projects onto a time like trajectory in the observer's space-time.
Thus, the directly observed dimensionality is always three plus one.
The effects of new coordinates must be inferred.  A similar problem occurs
in the transition to four dimensions.  Starting with
three space dimensions, the metric
assigns positions and distances to palpable objects.  The pythagorean
theorem can be verified (or falsified) by direct
measurement. Relativity demonstrates this
space to be incomplete.  It must be extended to include time.
In doing this, the developed procedure of using light beams
to find the components of the four metric, is not even
qualitatively equivalent to the use of a simple ruler.
In space-time, the experimental determination of geodesics is completely
different.   Presumably, any further extension beyond four dimensions
will involve crucial conceptual modifications.
By analogy, one would expect five dimensional measurements to be
even less pythagorean than four dimensional measurements.

Modifications in the metrical interpretation mitigate the need for
cylindricity or dynamical compactification.
Instead, the use of null five-vectors is sufficient.
Arbitrary non-null five vectors can be
constructed but they are not used here to represent physical
quantities.  As the geodesics of light are null in four-space,
the geodesics of particles are null in five-space.
This seems to be important for the relationship between Weyl and
Kaluza-Klein theories as well as for the proper inclusion of quantum
effects.  If the velocity vector is null, then the motion can be
represented by four parameters alone.
The essential vector length variations of Weyl theory are
incompatible with a Riemannian theory unless the vectors map into null
Riemannian vectors.  A fractional change in a Weyl vector can be mapped
onto a null vector because a fraction change in a null Riemannian
vector is permitted.   In this way, transformations within the
Riemannian system can be related to apparent conformal effects.
A non-null extension may be possible but
such vectors cannot be related to a Weyl theory.
A rigid metrical five dimensional structure that is analogous to
our perception of three-space is assumed in many other
theories.  These, even when
followed by the application of spontaneous compactification will not
allow for the implicit inclusion of quantum mechanics.

Because of the physics of geometrization, the construction
of lagrangians has not been found useful.
The physical quantities that have reasonable covariance properties are often
null or unavailable and the usual suppositions lead to quantities
which are identically zero.  Lagrangian mechanics was originally motivated
by the need to extend calculations
to situations involving constraints, such as rigid bodies or contact
forces.  This motivation is unjustified
in relativistic theories since rigid bodies do not exist and contact
elasticities are finite.  Furthermore, the
origin of these contact constraints is directly from the
quantum mechanics of collectively interacting
particles.  This questions the epistemological assignment of
lagrangians as a basis for quantum mechanics.  There is no apriori
reason to believe that theories developed
to describe  constraints, should provide a basis for microphysics.
The convenience of choosing interaction terms in a lagrangian increases
descriptive strength but reduces predictive power.  It has
been found necessary and advantageous to choose differential equations
without recourse to any construction of classical mechanics.

A few less controversial questions may be worth mentioning.
A full discussion of second quantization is not made here.
Such a formalism would represent
the description of more than one particle,
possibly including particle creation and annihilation.   These complexities
will have to be deferred until the one particle system is better understood.
At present, only a partial version of a geometrically appropriate
quantum electrodynamics is  available.   A complete
systematic discussion is needed.  Spin, weak and strong interactions
are also deferred.  An  understanding of
elementary quantum methods is deemed a necessary prerequisite.

A number of other constructions found in the literature have not been
incorporated.  Magnetic monopoles are not used.  Five dimensional monopole
solutions are supposed
nonphysical and may be mathematical ghosts~\cite{r31}.  Strings~\cite{r32}
may show  similarities but do no follow the formalism.
There are no discrete lattices.  Solitons are not used.
Space-time is not treated stochastically~\cite{r33}.  There is
no torsion~\cite{r34}.  There is no topological compactification,
dynamical or otherwise~\cite{r16}.
Because of the difficulty of precisely defining the difference between quantum
and non-quantum theories, classical mechanics is intended to refer to the
limit, as $\hbar \to 0$ and not to any fundamental classical theory.

These particular assumptions should provide a usable basic starting point.
Surely reality is more complicated.

\section{NEUTRAL FIVE-SPACE}
\label{nfs}

It is useful to associate the fifth dimension of a five dimensional
Riemannian space with the proper time.  Let the  defining equation
for $d\tau$ be written as:
\begin{equation}
\dot \gamma_{mn}dx^m dx^n = 0 \label{fdpte}
\end{equation}
where $dx^5 \equiv d\tau,\dot \gamma_{55}=-1,
\dot \gamma_{5\mu}=0$ and $\dot g_{\mu\nu} = \dot \gamma_{mn}.$
The Einstein summation is in effect. Lower case greek indices are summed
over four values and lower case latin indices are summed over five.
Equation~(\ref{fdpte}) is intended to apply to the standard observer of four
dimensional
space-time.  It is proposed that nullity of five displacements is
maintained for primitive particles even when off diagonal
terms are appended.

A neutral particle has a
fifth coordinate defined by the path integral of $d\tau=dx^5$
The space-time coordinates define $\tau$, up to an additive constant.
Further constructions will indicate that
electromagnetic  or quantum effects occur when $\gamma_{\mu 5} \neq 0$.
It is supposed that the nullity of displacements for charged neutral
particles is maintained in the  presence of the electromagnetic field.
The particle intrinsic fifth coordinate will not necessarily be the
same as the neutral space proper time.
In this case, an inferred physical comparison of
$\gamma_{\mu 5}, \gamma_{5 5},$ and $g_{\mu\nu}$,
can determine a physical value for
the new terms $\gamma_{m 5}$.
In the neutral case $\gamma_{55}$ is not well defined and is arbitrarily
set to -1.  Time like displacements are represented by real proper time values
in conventional units.

For classical relativity, the accumulation of differences in the
proper time along distinct trajectories is second order.  No cross
terms, $\gamma_{\mu 5}$, occur.
The twins of the twin paradox are neutral objects and experience second
order age corrections. The direct observation of first
order terms, which are only assigned to primitive particles,
is not possible because a real physical clock cannot be attached
to the five space trajectory.

\section{EQUIVALENCE FOR CHARGED QUANTUM PARTICLES}
\label{ecq}

The usual principle of equivalence is intended, historically,
 for uncharged
classical particles in a four dimensional theory~\cite{r13,r35}.
When attempting to extend this, the existence of particles with
different charge to mass ratios causes an essential difficulty.
Even more, quantum diffraction
depends inversely on the mass and destroys compositional additivity.
A principle of equivalence must therefore be constructed for a
single isolated particle characterized by particular charge and
mass value.

Particles having different interactions, (electromagnetic or quantum)
must use tensors of different internal construction.
The particle properties are incorporated into the
fields that determine the local particle rest frame.
The mass must enter with a factor $\hbar$.  As will be shown
in a  paper to follow, this factor is  a scale size of
the fifth dimension much as the speed of light
is  the scale size of the fourth dimension.
This construction presumes that the mass ratios of fundamental particles
are ultimately derivable geometrical quantities.  Factors
of $\hbar$ are taken to appear in combination with all masses.
This factor also appears in the fine structure
constant because the quantum manifestation of the inertial force is
a standard quantity against which the electric forces are measured.

The geometrical extension can be motivated by physical argument.
Consider a gedanken experiment performed on two idealized particles.
Each of these should be isolated, non-composite, stable, and
of finite mass.  The first one, denoted by $N$ is neutral
and does not respond to
electromagnetic fields.  The second, denoted by $C$ is charged.  It
responds in the classical limit
according to the Lorentz force law.  Following figure~(\ref{f1}),
the two particles begin on the left with coincident motion.
A single rest frame making both particles equivalent can be obtained.

The particles separate as they traverse a region where the electromagnetic
field tensor is nonzero.   The field in this region
can be adjusted so that after exiting, both particles converge to intersect at
space-time point $P_2$ with distinct velocities $U^\mu(C) \neq U^\mu(N)$.
At $P_2$, the particles cannot be made
equivalent even though they were equivalent at $P_1$.
The usual sense of the equivalence principle dictates that it should
be possible to perform continuous
coordinate transformations to find an invariant local rest frame for
either particle.  Within general relativity, there is no way to do this.

To resolve the paradox, additional geometrical quantities can be used to
describe velocities that may have electromagnetic (or quantum) origin.
Either a four dimensional non-Riemannian Weyl geometry or a five
dimensional Riemannian geometry is possible.
These supply additional electromagnetic-geometrical transformations
which describe the
motion of charged particles without affecting the motion
of idealized neutral objects. In this way, a principle of
equivalence can be used.

The nullity of the displacement vector, as applied to an individual particle
is related to the concept of a Killing vector.  It will
be seen that the equivalence of the trajectories under displacements
is related to the equivalence of the different forces that might
cause a deflection.  Different parts of the trajectory cannot be equivalent
as viewed by the neutral observer because the acceleration appears to
have different causative explanations.  That is, the effective field
and source currents vary along the congruence.

Several studies have been made of the intrinsic relationship of
the Aharonov-Bohm effect~\cite{r36} with five dimensional theories~\cite{r37}.
As a realizable example, applied to this gedanken experiment,
it is particularly interesting since
neither the neutral nor the charged particle
actually enters the region where the field tensor is non-zero.  There is
however a difference in deflection which is experimentally observable
and cannot be described by the conventional principle of equivalence
in space-time.

\section{FIXED GAUGE CLASSICAL THEORIES }
\label{fxg}

The metric tensor, $g_{\mu\nu}$, and electromagnetic vector potential,
$A_\mu$, have known classical geometrical interpretations. It is
desirable to adjoin a wave function without a first quantization
process.  As a first step,
the classical equations must be rewritten in a fixed gauge form
{}~\cite{r38,r19,r21}.  The usual gauge freedom is assumed non-fundamental
and is transformed to a fixed value specified by making  the action
identically zero.  (Any analogous  wave function will have its phase
removed.)
The standard classical
five metric, having signature $(+,-,-,-,-)$ is appropriate
and can be written in the form
\begin{equation}
\gamma_{mn}=\pmatrix{g_{\mu\nu}- \A_\mu \A_\nu & \A_\mu \cr
\A_\nu & -1 \cr}. \label{fdm}
\end{equation}
where $\A_\mu={e \over m} A_\mu$.
It is sufficient to show that a velocity defined by
\begin{equation}
{dx^\mu \over dw}=\A^\mu \label{nfxf}
\end{equation}
defines a classical null geodesic. The absolute derivative of the above is
\begin{equation}
{ {d^2x^\mu \over dw^2}+\dot {\mu \brace \nu \beta}
{dx^\nu\over dw}{dx^\beta \over dw}=
\dot g^{\mu\beta} \left[ {\partial \A_\beta \over \partial x^\lambda}
{dx^\lambda \over dw}- \dot {\lambda \brace \nu \beta}
\A_\lambda{dx^\nu \over dw}\right] }
\end{equation}
wherein the Christoffel symbols are calculated with respect to the
neutral space four metric $\dot g_{\mu\nu}$
This can be converted into the conventional form since the Hamilton-Jacobi
equation with fixed gauges is ${\A_\mu \A_\nu \dot g^{\mu\nu}=1}$.
The covariant derivative of this equation can be used to eliminate the
second Christoffel symbol which gives
\begin{equation}
{{d^2x^\mu \over dw^2}}
+ {\dot {\mu \brace \nu \beta}
{dx^\nu \over dw}{dx^\beta \over dw}} =
{ \dot g^{\mu\beta} \left(
{\partial \A_\beta \over \partial x^\nu}-
{\partial \A_\nu \over \partial x^\beta}
\right) {dx^\nu \over dw}} \label{sqq}.
\end{equation}
Here, the path parameter ${w}$ is normalized relative
to the neutral space four metric.  Direct derivations from the
five metric are given in references~\cite{r7},
either by the Lagrangian method or by
parallel transport. These agree with the above fixed gauge
calculation and give the
the combined force law.
This establishes that the fixed gauge equation
of motion~(\ref{nfxf})
is a valid construction in the presence of gravitational and
electromagnetic fields.

This description defines a congruence
of motion which is the set of integral curves of a fixed gauge
vector potential field.  Such an individual congruence, which is
a general solution of the Hamilton-Jacobi system,
can be developed into a quantum state.  Selection of a whole congruence
need not specify
a particular trajectory as the unique trajectory of a particle. To
completely specify
such a quantum state, a quantum field equation is required.
For a scalar field, the Klein-Gordon equation is used.
Such a quantum solution, when written
in a fixed gauge representation cannot in general have a normalized
vector potential as in the classical  representation.  The magnitude
$ A_\mu A_\nu g^{\mu \nu}$ is not constant.   This is true
even though the unnormalized vector potential still satisfies
the inhomogeneous Maxwell equations.
To make the system amenable to quantization, it is assumed at this point
that the unnormalized vector potential is still suitable as a purveyor
of the physical quantity within the geometrical system.  This identification
can be applied to both the Weyl and Kaluza theories.

The resulting congruence contains, under reasonable conditions and with
reasonable physical assumptions, the information that was contained in
the original wave function.   Any phase information
has been incorporated directly into $A_\nu$, only the magnitude of
$\psi$  need be regenerated.
This can often be done, in practical
cases from the current conservation alone.  Global reconstruction
may require recourse to the field equation and boundary conditions.
The essential idea is that such a congruence is sufficient to represent
the essential part of a single particle quantum state function.

\section{LOCAL FIVE TRANSFORMATIONS }
\label{fdt}

The local structure of the five dimensional coordinate system is
larger than the  the four dimensional Lorentz group.  The additional
transformations are necessary  for the principle of equivalence
discussed in section~\ref{ecq}.   An interpretation can be developed
by applying the new
transformations as local point operations on a charged particle.
A related problem is mentioned in ~\cite{r39}.

The usual transformations involving the observer's coordinates
$x_\mu$ still apply and affect the vector potential covariantly.
The transformations involving $\tau$ are new, and
because of the fixed gauge assumption, have specific physical
effects and interpretations.  They are affine and orthogonality
is not maintained.
The most useful of these is
\begin{equation}
x^{\mu \prime}=x^\mu+\A^\mu\tau \label{alct}
\end{equation}
and changes the four-velocity of the charged particle according to
\begin{equation}
{dx^{\mu \prime}\over ds}={dx^\mu \over ds}+\A^\mu{d\tau \over ds}
\end{equation}
The electromagnetic vector potential is thereby identified
with a shear transformation.
If the arbitrary path increment $ds$ is selected equal to the apparent
proper time change, $dw$, the
transformation~(\ref{alct}) gives a new velocity for the particle.
Such transformations represent a realignment of the five dimensional
space. It is used to describe velocity changes
that are of quantum or electromagnetic origin as distinguished
from those that are gravitational.

Using $p^\mu =mv^\mu$ for the kinetic momentum,
\begin{equation}
p_\mu=p_\mu^\prime-m\A_\mu
\end{equation}
identifies the quantum nature of the transformation because
of the similarity with the minimal substitution.
This assumption
identifies the motion as a first order quantity relative to the
neutral system.  A Lorentz transformation, as a contact transformation
in space-time, changes the velocity  of the particle
and also the value of $\A_\mu$.   The new value represents
the instantaneous particle motion and also the local effective transformed
orientation of the five dimensional space.   The electromagnetic
effects are not modeled by Lorentz group elements but by the
new aspects of the five dimensional coordinate system.

The gedanken experiment of section~(\ref{ecq}) supports
this interpretation.  Within the interaction region
of figure~(\ref{f1}), five dimensional
effects occur which cause a deflection of the particle relative
to the unvarying neutral geodesics.  Once free from
the interaction, and arriving at $P_2$, the integrated effects of the
electromagnetic field leave a residual velocity $U_2(C)$ that is
distinct in ontology from the neutral velocity $U_2(N)$.   This is
implicitly defined by the orientation of the five dimensional system,
which has an internal geometrical distortion that is not represented in
neutral space-time.  This integrated effect persists outside of the
interaction region.   It can, in fact, describe Aharonov-Bohm
type effects.  The velocity of the charged particle cannot be reduced
to the velocity of the neutral particle by a space-time
Lorentz transformation. A similar discussion can be made for a fixed gauge
Weyl theory.  The charge particle motion is varied by the non-Riemannian
part of the connections which are outside the range of space-time
Lorentz transformations.

\section{INERTIA AND MEASUREMENT}
\label{gin}

An important property of this description of a quantum state is
that there is no formal identification of the mass.
Moreover, various different field values $A_\mu$ can correspond
to the motion of a particle having
possibly continuously varying value.  There is no
inertial frame because arbitrary transformations are allowed.
The motion is inertia free.  This mathematical concept of
motion is important to allow the resolution of the very different
concepts of inertia that prevail for the interactions studied.
The relation between the physical inertia and its
numerical representation  is not the same for
classical, quantum, and gravitational theories.
For gravity, inertial effects are independent of the value
of the mass, as long as it is nonzero.   Electromagnetism
has a characteristic $e/m$ value and quantum mechanics scales
inversely.

For conventional derivations of quantum mechanics, the inertial properties
of particles are assumed to be intrinsic and are eventually introduced
through the classical hamiltonian or lagrangian.
This, prevents the identification of the mass with the field equations.
This epistemology must be changed
in a geometrical construction.  The inertia is fundamental
and originates
with the quantum processes of diffraction and interference.
These effects  must  come from the quantum geometrical
system without the intervention of classical physics.

Measurement of the inertia of a particle must be considered
fundamentally quantum.  The classical observation is the
good fortune of a simple experiment.  It is in reality a measure of the ratio
of the Compton wavelength of the test particle to one or more Compton
wavelengths in the measuring apparatus.
In the classical limit, the diffraction effects become negligible
and inertia, as is classically understood, remains.

The observation of inertia depends on having
quantum stabilized dimensional standards.
The observers coordinate system must be calibrated in a consistent, systematic
manner~\cite{r41}. Even the elementary demonstration of
Newton using a bucket of water or spheres connected by a cord refers
to sized objects.  A single length standard is represented by
a standard clock that is quantum based and
measures in terms of $\hbar / mc^2$.   A bouncing light
beam clock requires that the mirrors be held at an absolute space-like
distance.  An apparatus to hold the mirrors must be built with atoms, or
other quantum objects of constant size determined equivalently by
values of ${\hbar / mc}$.

The only alternative time scale is one
based on the Planck length. The relationship is addressed in the
work of Dirac and others on large numbers~\cite{r42,r15}.
It is supposed here that the Planck length may be cosmological rather than
fundamental.  The relationship of the Planck clock to the atomic clock
remains a goal of theoretical and experimental research~\cite{r43}.
It is assumed that the fine structure constant $\alpha$
is an invariant and that any of the
time scales which can be formed by $\alpha^k({\hbar / m})$ are the same.  For
such situations as $\alpha$ might be allowed to vary, ${\hbar / m}$
is taken as the fundamental quantity.  Other interactions, weak and
strong, are presumed unessential to chronometry. Figure~(\ref{f2}) shows
a quantum clock that avoids these problems and provides a theoretical time
standard.

Because the mass is taken out of the classical arena,
the phenomenology of spacelike measurements must be carefully
reconsidered.  The accepted construction is shown in figure~(\ref{f3}).  Two
particles intersect two spacelike separated points $x_1$ and $x_2$.  A
light beam is sent from particle $1$ at time $t_A$ to particle $2$ so that
it arrives at $t=0$. The beam is immediately returned to particle $1$ arriving
at time $t_B$. The distance $x_1-x_2$ is to be inferred from the measured
delays $t_A$ and $t_B$.  If, however, the particle $1$ is described by
quantum mechanics, conventional theory proposes that the trajectory is
ambiguously defined if it is defined at all.  It may in fact be thought
to intersect the  $x$ axis anywhere between $x_2$ and $2x_1-x_2$, which are
the relativistically allowed limits.  The distance $x_2-x_1$ can no longer be
measured. This is a serious metaphysical failure.  Even a practical analysis
shows that the observational
determination of a macroscopic observer's metric for a particular coordinate
system can only be done if the experimental system is sufficiently large to be
treated classically~\cite{r29}.  Because of this failure, the classical
construction must be replaced if a fundamental sense of space-time
is to be maintained.   The assumed classical geodesy must be considered a
phenomenological result.

Following Mach~\cite{r40}, attempts have been made to associate inertia with
gravitational forces by the action of distant objects~\cite{r44}.
{}From the second order calculation made by Thirring~\cite{r45}, it
is known that there is an effect on inertia. Because it is not possible
to manipulate the distant objects experimentally,
the choice between using a relative concept of inertia and an absolute
concept is probably to be made by mathematical convenience.  The
inertia free description of the five dimensional system can
accomplish this.
The usual absence of first order structure is remedied
by the fixed gauge construction.  The boundary conditions
of Mach may then be supplied by the standard quantum (or Hamilton-Jacobi)
boundary conditions.  The resulting fixed gauge system
has the mathematical capacity to incorporate a fundamental
construction of inertia.

With three fields, there will be three corresponding
sets of boundary conditions. Electromagnetic radiation(and the implied
acceleration) is to be related to boundary conditions defined by an
external absorber. Gravitational radiation is analogous but more
complicated.
And for quantum mechanics, the sense of inertia implied by the eigenstates
is determined by the space-time boundary conditions.
The separation of these effects may depend on the particular
choice of a neutral physical space-time.
The quantum aspects are the more important lowest order effects.
Electromagnetic effects, when separated from quantum effects,
are first order and gravitational forces are second order.  This supports
generally the modern ideas that relate inertia to electromagnetic fields.

The study of the properties of a quantum congruence shows the
importance of quantum boundary conditions for the expression of inertia.
Consider, as sketched in figure~(\ref{f4}), an experiment
consisting of particles that traverse a series of unaligned apertures.
A certain fraction will diffract through
each hole in turn and finally be detected.  These
cannot be said to move on a classical
trajectory.  Moreover, the number of apertures
can be increased indefinitely as
long as the they are not too small and the particles are not required
to have unattainable velocities.  Otherwise, the trajectories are
nearly arbitrary.  The loss of particle counts is not important and is
actually caused by implied
absorptive boundary conditions.  Rather than have the particles hit the
screens, each aperture could be connected to the previous one by
a small tube upon which the wave function is set to zero.  In either
case, the motion is convoluted but does not require
external classical fields.   If $\A_\mu(x,t)$ be chosen everywhere
tangent to the probability current density, the
five dimensional fixed gauge congruence can describe this motion.
By adjusting the five dimensional ``cut'' coordinate
transformation to align the vector potential
with the local current, a quantum particle can be followed as it
traverses the experiment.  The five dimensional geometry
is sufficient to describe elementary quantum motion.

Fundamental quantum inertia is
the constraint put on the trajectories by the geometry
as external forces or boundary conditions are applied through
the field equations.
These should be constructed from invariants~\cite{r19,r21,r50,r1}.
All of this requires more formalism.
Because the trajectories of charged particles are not straight lines
with respect to a neutral frame,  curvilinear coordinates are
required.  This elementary picture, that uses straight trajectory
segments must be expanded into a full structure capable of
describing complex motion.

\section{THE TRANSITION TO FIVE DIMENSIONS}
\label{tfd}

The particle congruence, as a mathematical representation of
a physical object, admits five dimensional coordinate
transformations analogous
with the four dimensional Lorentz transformations.
These can be used to construct and interpret five dimensional metric tensors.
To first order, a differential vector $(dx^\mu,d\tau)$, in a coordinate
system $(x^\mu,\tau)$,
will be mapped linearly onto a differential vector
$(dx^{\mu\prime},d\tau^\prime)$ in a coordinate system
$(x^{\mu\prime},\tau^\prime)$.
In this approximation,
\begin{equation}
dx^{\mu\prime}=C_5dx^\mu+C^\mu d\tau
\end{equation}
\begin{equation}
d\tau^\prime=D_5 d\tau+D_\mu dx^\mu
\end{equation}
where $C$ and $D$ must satisfy integrability conditions.

The coefficient $C_5$ describes a local conformal transformation
for space-time.  If $C_5$ is constrained to unity,
conformal effects are removed from the coordinate transformation.  The case
$C^\mu \neq 0$ is useful to
eliminate $\tau$ dependence when it is present.
The coefficient $D_5$ may be at most indirectly observed because it
involves the normalization of $\tau$ over extended space-time.
The remaining coefficient $D_\mu$ produces quantum and
electromagnetic effects.
The invariance properties of this term have been studied by Klein
and others~\cite{r57}.  It is usually identified as an invariant
gauge transformation.  In a fixed gauge theory, it operates
on the particle state and changes the velocity, mass, and
effective wave function.  It is an integral part of
the extended principle of equivalence.

This last term is best studied separately
by examining the shear transformation.
\begin{equation}
x^{\mu\prime}=x^\mu
\end{equation}
\begin{equation}
\tau^\prime=\Phi(x^\mu)+\tau
\end{equation}
which becomes
\begin{equation}
dx^{\mu \prime} =dx^\mu \label{cutt1}
\end{equation}
\begin{equation}
d\tau^\prime = \Phi_{\mu} dx^\mu +d \tau \label{cutt2}
\end{equation}
and where the coefficient $\Phi_\mu \equiv \Phi_{,\mu}$ must always be exact.

The transformed five metric is calculated directly from
the invariance of the line element.
\begin{equation}
dx^{m\prime}dx^{n\prime}\gamma_{mn}^\prime=
dx^{\mu}dx^{\nu}g_{\mu\nu}-d\tau^2=
dx^{\mu\prime}dx^{\nu\prime}\dot g_{\mu\nu}
-(d\tau-\Phi_{,\mu} dx^{\mu\prime})^2
\end{equation}
giving
\begin{equation}
{\gamma_{mn}=\pmatrix{ g^{\mu\nu}-\Phi_{,\mu} \Phi_{,\nu} &
\Phi_{,\mu} \cr \Phi_{,\nu} & -1 \cr}}.
\end{equation}

The local five-Lorentz transformations can also be calculated from
\begin{equation}
dx^{a\prime}=\Lambda^a_{\cdot b}dx^b
\end{equation}
which gives
\begin{equation}
{\Lambda^a_{\cdot b}=\pmatrix{ \delta^\alpha_\beta &
0 \cr \Phi_{,\beta} & 1 \cr}} \label{lmctv}
\end{equation}
and again demonstrates the shear character.

In five dimensions, two gauge coefficients may be present rather
than the single gauge factor $\lambda$ of the Weyl
theory.  To make the transition from neutral space, it is necessary to write,
in the most general case,
\begin{equation}
d\tau^2\chi^2=\lambda \dot g_{\mu\nu} dx^\mu dx^\nu
\end{equation}
where $\chi$ and $\lambda$ can both be functions of position.
The ratio $\lambda /\chi^2$ can no longer be treated
as one gauge function.  When $\Phi_\mu \neq 0$, the factors have
distinct effects since there are off diagonal
terms in $\gamma_{ab}$.
If such a conformal transformation is applied as a point
transformation it changes the physical fields.
In addition there is possibly an overall conformal multiplier $\omega$.
These factors do not affect the projection of the congruence
onto space-time. They must either be invariance transformations
or else must change the identities of the physical fields that
accompany the motion.

The application of the conformal transformations along
with the extended Lorentz
transformations generates a more
complicated single particle metric from neutral space.
\begin{equation}
{\gamma_{mn}^\prime=
\pmatrix{ (\lambda \dot g_{\mu\nu}-\Phi_{,\mu} \Phi_{,\nu})\omega &
{\omega\Phi_{,\mu} / \chi} \cr
{\omega\Phi_{,\nu} / \chi} & -{\omega / \chi^2} \cr}}
\end{equation}
Setting $\omega=\chi^2$ and $\lambda^\prime = \lambda \omega$ with
$\A_\mu=\Phi_\mu \chi $ this metric becomes
\begin{equation}
{\gamma_{mn}^\prime=
\pmatrix{\lambda^\prime \dot g_{\mu\nu}-\A_\mu \A_\nu &
\A_\mu  \cr \A_\nu & -1 \cr}} \label{gdkkf}
\end{equation}
The new quantity $\A_\mu=\chi\Phi_\mu$ is not in general integrable and
can be associated with the electromagnetic field.

This metric is a generalization of the standard Kaluza-Klein form.
The conformal transformations have been joined with the coordinate
transformations to allow the generation of quantum and electromagnetic
effects.
The resulting five metric, expressed in terms of fixed gauge
quantities, is distinct from the conventional Kaluza-Klein theory because
it represents the microphysics of
a single quantum particle.    It is a quantum object, not
by any process of quantization, but by the fixed gauge assumption and
the inherent quantum nature of geometry.

Because these conformal factors do not change the direction of $\A_\mu$,
the five dimensional null geodesics of section~\ref{fxg}
generalize immediately to the same trajectories proposed by
the quantum-Weyl theory~\cite{r19}.
External interactions including the electromagnetic source terms
presumably
can be characterized by combinations of the conformal factors
$\chi$, $\lambda$, and $\omega$.  It is the transformation
of source terms implied by changes in conformal factors
that is an essential part of extended equivalence.  By setting
different conformal factors (as observed from the neutral space)
different mechanistic combinations of electromagnetic,
quantum and gravitational effects can be ascribed to a given congruence.
The subtleties of the existence of a Killing vector are now more
apparent.  Each point of the congruence is equivalent as far as
simple motion is concerned.  If however, the particular metric coefficients
(that are the expressed determining influence of that motion) are included,
each point can have different local external fields.

When constructing an arbitrary vector potential from a gradient,
a single integrating factor
may not always exist.  In this case, a more complex approach to
interaction is necessary.
A complete physical determination awaits a set of
quantum-Einstein-Maxwell source equations.  For a single external source
particle, the
integrating factor can be found in the rest frame of the source.  For
multiple source particles, additivity fails and a more involved
analysis is needed.  It is conjectured that the requirement
of multiple integrating factors is naturally satisfied by the use of
multiple source particles.

\section{QUANTUM GEODESICS}
\label{qgd}

The geodesic system of section~(\ref{fxg}), describing classical
motion, can be applied directly to the  quantum case.  From
standard  theory, it is known that
the Klein-Gordon conserved current in fixed gauge form is
\begin{equation}
{P^\mu={e \over m} \psi^\ast\psi A^\mu }
\end{equation}
This defines a congruence for a solution of the quantum field equation.
It is to be associated with one or more 5-metrics.
The trajectories are to be defined by a first order fixed gauge
equation, in form identical to the classical case.
The changing length of $\A_\mu$ is now essential
to the quantum and electrodynamic observations.

A modified normalization can be defined by
${A_\mu^\star=\xi A_\mu }$.
Let the factor ${\xi}$ be chosen so that
${{e^2 / m^2 }A_\nu^\star A_\mu^\star \dot g^{\mu\nu}=1}$.
This starred vector potential can be used to define a trajectory
with a parameter ${w}$ that is
entirely analogous to the classical case except that
$A_\mu^\star$ is not a solution of Maxwell's equations.
\begin{equation}
{{dx^\mu \over dw}={e \over m}A^{\star\mu} \equiv
{e \over m}A_\nu^\star \dot g^{\nu\mu}} \label{qntgeo}
\end{equation}
The $dx^5$ dependence can be chosen to keep the five displacement null.

{}From the arguments  of section~(\ref{tfd}) it is expected that
an appropriate five dimensional metric is of the form
\begin{equation}
{\gamma_{mn}=\omega
\pmatrix{g_{\mu\nu}- \A_\mu \A_\nu
 & \A_\mu \cr \A_\nu & -1 \cr}}
\end{equation}
where ${g_{\mu\nu}}$ and ${\A_\mu}$ are
fixed gauge quantum fields.
To show that these are geodesics,
a coordinate transformation can be defined by
${dx^{5 \prime} =\xi dx^5}$.
This choice depends on the particular congruence.
It is not integrable generally
but can be chosen uniquely by executing the integration along
the congruence.

Choosing $\omega=\xi^2\omega^\prime$, the five dimensional metric now becomes
\begin{equation}
\gamma_{mn} =
\pmatrix{\xi^2g_{\mu\nu}-
\A^\star_\mu \A^\star_\mu & \A^\star_\mu \cr
\A^\star_\nu & -1 \cr}
\end{equation}
and has the same form as the classical theories.
It must have geodesics as given by~(\ref{qntgeo}).  These are quantum geodesics
because the motion they predict gives a correct description of
statistical measurements of quantum states in combined gravitational and
electromagnetic fields.
They are tangent to the accepted probability density current
in space-time.  By the argument of
section~(\ref{fxg}), the observed
second order equation can be found by taking the
absolute derivative of the quantum trajectory
relative to the observers' metric ${\dot g_{\mu\nu}}$.
This gives
\begin{equation}
{ {d^2x^\mu \over dw^2}+
\dot {\mu \brace \epsilon \lambda}
{dx^\epsilon \over dw }
{dx^\lambda  \over dw }=
{e \over m}\dot g^{\mu\beta} \left(
{\partial A^\star_\beta \over \partial x^\lambda}-
{\partial A^\star_\lambda \over \partial x^\beta}
\right){dx^\lambda \over dw} }.
\end{equation}

Quantum forces are included  by way of the rescaled vector potential.
The particle motion is always tangent to $\A_\mu$ but the second order
derivatives are attributed to different physical fields by the measuring
process.
Resubstituting ${A^\star_\mu=\xi A_\mu}$,
\begin{eqnarray}
{d^2x^\mu \over dw^2} + \dot {\mu \brace \epsilon \lambda}
{dx^\epsilon \over dw}{dx^\lambda  \over dw}=
{e \over m} \dot g^{\mu\beta} \left[
{\partial (\xi A_\beta) \over \partial x^\lambda}-
{\partial (\xi A_\lambda) \over \partial x^\beta}\right] = \nonumber \\
{e \over m} \dot g^{\mu\lambda} \left(
{\partial A_\beta \over \partial x^\lambda}-
{\partial A_\lambda \over \partial x^\beta}\right) {dx^\lambda \over dw}+
\dot g^{\mu\beta}(\xi_\lambda A_\nu - \xi_\beta A_\lambda)
{dx^\lambda \over dw}+
{e \over m} \dot g^{\mu\beta}(\xi-1) \left(
{\partial A_\beta \over \partial x^\lambda}-
{\partial A_\lambda \over \partial x^\beta} \right){dx^\lambda \over dw}.
\end{eqnarray}
The first term on the left is the acceleration of the particle and
the second includes all classical
gravitational and fictious (inertial) forces.
The right hand side splits into three parts. The first is the conventional
electromagnetic force.  The second and third are quantum forces which can be
named respectively convected and parametric because of the dependence on
the factor $\xi$.  The representation is of the motion of a single quantum
particle interacting through fields alone.  With this formalism, it is
easy to see that an electron diffraction pattern
should be distorted smoothly by an external field, either electromagnetic
or gravitational.

Much like the null four geodesics of light beams,
the null five geodesics do not change under an overall conformal
transformation.  The first order defining equation for a quantum geodesic
can be written as
\begin{equation}
{dx^\mu \over ds}=g^{\mu\nu}\A_\nu
\end{equation}
where $ds$ is not in general affine.  The above result still holds
because the derivative of the projected
trajectory is made by the observer in terms of the neutral space metric.
The path parameter $w$, normalized by $\dot g_{\mu\nu}$, is used
to display the apparent acceleration of the inferred four-velocity of
the particle.

The overall conformal invariance can be used to provide an
additional demonstration that the probability current
is directed along null five geodesics.
One can consider a particular factor ${\zeta}$ such that it is
the integrating factor of the electromagnetic vector potential.
As before let ${\A_\mu \zeta = \Phi_{,\mu}}$ for some scalar $\Phi$ and
factor $\zeta$.
In addition, the fifth coordinate
can be transformed by ${dx^5=\zeta dx^5}$ using again the
integration along the congruence.  The metric after the transformation is
\begin{equation}
{\gamma_{mn}=\pmatrix{\zeta^2 g_{\mu\nu}-\Phi_{,\mu} \Phi_{,\nu} &
\Phi_{,\mu} \cr \Phi_{,\nu} & -1 \cr}}
\end{equation}
and gives after the cut transformation
${x^{5\prime}=x^5 -\Phi}$
the diagonal form
\begin{equation}
{\gamma_{mn}=\pmatrix{\zeta^2 g_{\mu\nu} &
0 \cr 0 & -1 \cr}}
\end{equation}

Again, more than one conformal-coordinate transformation may
be needed to integrate the vector potential.
Geodesics of this diagonal metric include the
point solutions $x^m=0$ that satisfy $dx^\mu / ds =0$. These
transform to the initial geodesics by the inverse of the above
sequence.  The residual factor ${\zeta^2 g_{\mu\nu}}$
contains information on interactions and probability densities.
In this coordinate system, each geodesic is represented by
one point which is apparently the hidden variable.
This point can be fixed at a certain
value of the proper time relative to an arbitrary initial surface.
The entire past and future history
of the particle congruence
can be calculated from the five metric by integrating
along geodesics in five-space.  For given determining
external fields, the intersection point of this trajectory with
a particular space like surface can
be found.  Each possible particle position measurement
can be thought to result from one geodesic in the collected congruence.
The congruence as a whole represents the motion generated by
the specific given quantum state.

\section{ON VON NEUMANN'S THEOREM}
\label{vnt}

A few comments may clarify the issue of hidden variables.  It is
necessary because in a fundamentally geometric theory,
it may be essential to describe particles with geodesics.
Such specific identification of point particle motion is adverse to
conventional interpretations of quantum mechanics.
A complete critical
discussion of the standard objections to hidden variable theories is outside
the scope of this paper.  When precisely applied, the usual objections are
valid for many alternate  theories.  Some of these alternate theories may also
disagree with experiment or may have some mathematical
error. Overall,  there is no generally accepted way of avoiding the
objections to  hidden variables.  For the purposes of this article
it is sufficient to point out a class
of geometrical theories that are not subject to the accepted objections.

One of the most serious practical problems
is that an assumption of the existence of hidden variables provides no
guidance as to how to find them.  Under these circumstances, the
question of definition is most important.
It is usually supposed that any reasonable
choice would be derived from the classical theory through the system of
observables constructed by Von-Neumann~\cite{r58}.
This is not the case here.  The mathematics of differential geometry
substitutes an inequivalent fundamental structure.
Since the quantum system is not to be derived from classical physics,
the use of differential operators as physical concepts is not required.
For the geometrical theories of this paper, classical mechanics is
fully rejected as a fundamental theory.
The conditions of Von Neumann's theorem fail because the hidden
variables no longer have to be defined according to his prescription.

While de Broglie's hidden variable concept,~\cite{r60}, is still
popular in the minds of many physicists~\cite{r61}, his
use of the operator substitution method allows
Von Neumann's argument to prevail.
This suppressive theorem can be applied when the derivative as a representative
of the physical momentum is presumed to  be a legal substitution
into the equations of classical mechanics.

Without bare physical operators,
all physical quantities commute, and Von Neumann's
theorem is avoided.  The concept of the classical
canonical momentum matched with the analogous operator oriented
quantum concept fail together.  There
is no quantum measurement theory in the usual sense.  It is worth noting
that if a quantum theory is defined
by a system of differential equations alone,
then there is no unambiguous way to define the canonical momentum.
All classical theory is avoided.

The process of operator substitution has no mathematical
precedent and no accepted mathematical justification.  The metaphysics of
quantization fails and the arguments of Von Neumann may not hold.
First quantization, in
actuality, serves the purpose of recreating
differential factors that are neglected in the phenomenological
perception of classical mechanics.  It is an essential method
in an historical context.  All of the accepted
derivations,~\cite{r59}, of the Klein-Gordon
equation, except possibly the one by Klein,  proceed along these lines.

Thus the geometrical view presented here is fundamentally different.
It is not a reformulation and new results are possible.
In particular,
the capability to address gravitational problems is added over
the conventional wisdom.  The rejection of first quantization may have effects
that extend beyond immediate considerations.  The repeated
circular derivation of quantum from classical and then classical from
quantum is implicit in much of modern physics.  This cyclic logic should be
broken at the point of first quantization.
Geometrical theories or any other new theory may need to be evaluated
against experiment and not current phenomenology.

\section{ON BELL'S THEOREM}
\label{obt}

Modern criticism of alternative quantum theories and hidden variables
usually is presented in the context of Bell's theorem~\cite{r62}.
Comparisons of theory and experiment have been found
to support the conventional formalism.  A number of measured
results~\cite{r63} lie outside the predictions of a large
class of alternate theories.
Bell's theorem is motivated partly from the work of
Von Neumann. The accepted quantum discussion depends on the use of
operators as physical quantities and
neglects relativistic effects.
A careful consideration of the assumptions
and arguments of Bell's theorem shows under what conditions
alternate constructions
might be mathematically possible and experimentally acceptable.

A serious weakness of Bell's analysis
centers around the approximations that are necessary to make a manifestly
relativistic or covariant theory non-relativistic.
When applied to alternate theories, Bell's objections
depend on the presence of an implied classical basis.
This has subtle implications that become part of the
quantum interpretation.
Moreover, the wave function is hypothesized without any predecessor.
Better that the wave function should be part of a geometrical space
and not an esoteric representation of some sort of fundamental
statistical essence.  In addition, some of the classical properties
appended during quantization, especially those relating to causality,
are misleading and inappropriate.

Radiation is manifestly relativistic and the mathematics and
sense of causality
 that deal with it
should be covariant.    The justification of the use of a
nonrelativistic limit for radiation comes only from
classical physics.   It is this metaphysics of motion in a relativistic
field that causes trouble.   It is known that
Bell's theorem cannot be applied to a theory like
quantum electrodynamics wherein advanced potentials are used.
Such fields, while undesirable in a classical theory, are
not forbidden by any concept of fundamental geometry.
A fully covariant geometrical theory
in which advanced fields are implicit, may be capable of predicting the
non-local results typical of photon correlation experiments.  Conversely
it is not possible to derive a non-relativistic quantum mechanics
from either quantum electrodynamics or geometrical theory without serious
conceptual compromises.

The implications of such classically imposed
causality are subtle.  For a fully microscopically deterministic theory,
an experiment with {\it complete} arbitrary initial conditions
(in the sense of the Einstein, Rosen, and Podolsky~\cite{r64}) is not apriori
possible.   It is believed that almost any experiment can be done,
but only a the small number of initial conditions allowed by
the current state of the universe are possible.  The
temperature of the experiment must be greater than the ambient noise.
Any experiment, because of electromagnetic radiation and distant gravitational
interaction may connect with
particles at the farthest reaches, past and future.
Our sense of causality is a large scale observation of these events.
A microscopic electromagnetic field
that does or does not go backwards in time can only be included or excluded
in so far as it does or does not explain scientific experience.
Moreover, since the microscopic structure of quantum
mechanics does not have an intrinsic direction,
formal symmetry between advanced and retarded potentials must be possible.

For a photon correlation experiment, the calculation using the
propagators of quantum electrodynamics has essential terms with advanced
dependency.  Because of implicit instantaneous interactions, non-relativistic
quantum mechanics implies the use of such fields but also { \it represses
any explicit reference to them}.  Since Bell's theorem
requires that all fields be retarded, the conclusions of this
theorem are avoided for quantum systems.  Even an implicit advanced
interaction voids Bell's theorem.
It is still unknown, whether in a more complete theory,
the presence of advanced potentials must be explicitly
displayed or whether they are an artifact of the mathematical methods.
Because of the demonstration by experiment, such interactions
must be integrated into the theory despite the counter-intuitive
indications of classical physics.

It is easy to show that the common geometrical theories already contain
advanced potentials.  Since a congruence of motion is to be
well defined, it must represent all physical effects.
In particular, the force of radiative reaction must be included.
This must be introduced into the vector potential by the
advanced potentials of other interacting particles.
The final overall prediction need not be interpreted
as an advanced propagation of information or energy.  A fixed gauge
system that includes all electromagnetic forces must be exempt from
Bell's theorem.

The relationship of geometrical theories to these
accepted experiments is important.
Most measurements use photon correlations.
For the present development of the geometrical description there is
no reason to consider calculations which are not equivalent to quantum
electrodynamics.  The geometrically preferred scheme is to use time symmetric
potentials without free fields.  Theoretical
 predictions of this type have already been
shown to agree with the standard versions~\cite{r25}.

The experiment by Aspect is representative of this class and is
straightforward to analyze.   A limitation to retarded
potentials is not required.  For a two photon correlated emission,
the emitting atom cannot execute the state transition without the
presence of the simultaneous advanced electrodynamic fields
from two or more absorbing
particles~\cite{r65}.   The correlation can be observed
only if the experiment is arranged so that both emitted photons are
collected in the experiment rather than in other parts of the universe.

There are also correlation experiments that involve particles.  In the
case where the interaction between the particles is mediated by
electromagnetic fields, the elementary geometrical theory should be
sufficient.  The nuclear experiments that use spin polarization~\cite{r66}
give important results for weak and strong interactions.
Formal predictions for these fields are not possible because the
geometrical structure used here is insufficient.  Physically, though
it is possible to argue that the results are not incongruent.  The
use of advanced potentials for other fields should
be adequate to predict the observed result.  In particular, the effects
of spin during scattering do not seem
to be time asymmetric.  These results should
be explainable using the nuclear equivalent of time symmetric
potentials.  The principle of equivalence is expected to have further
extensions so that  even spin couplings are  replaced by the effects of
covariant geometrical fields.

The entanglement of multiparticle
states is assumed to be the integrated effect of
the physical interaction that causes the state to form.   The
entangled states described early by Schr\"odinger,~\cite{r67},
that are so characteristic
of quantum mechanics are interpreted as a persistent
geometrical distortion that is
inseparable from the interaction that connects the particles.

Time symmetric electrodynamics, developed as a classical
theory~\cite{r68,r22} must be modified to allow for
quantum effects.  The macroscopic classical approximation must be
the limit of the quantum mechanical transitions that really occur.
The use of the classical theory as the metaphysical
precursor of the quantum field leads to confusion.
This sort of quantum electrodynamics has been applied to
cosmological situations~\cite{r69}.  Therein, the quantization
of time symmetric electrodynamics,~\cite{r70} is based on classical
physics.   The unusual predicted results with respect to absorption
and emission seem
to be due to the assumptions of those classical properties before the
quantization is accomplished.  The confusion
is related to the similarity of classical electrodynamics to possible
primitive geometrical forms of electrodynamics.  A precise notion of
classical absorption or emission cannot be defined until the fundamental
processes of discrete quantum  absorption or emission are understood.
More recent experiments on quantum electrodynamics support the
concept that
the pre-quantization assumption of absorption and emission
is not be justified~\cite{r71}.

\section{DISCUSSION}
\label{disc}

A number of important issues are raised by the application of covariant
geodesics to quantum particles.  This possibility is not
part of the accepted formulation of quantum theory.  It is of some
interest because the fundamental  construction is simpler than
other quantum gravitational theories.
The real issue is whether the use of such a construction can produce the
essential results and complex phenomenology of modern quantum experiments.
This article specifies how to begin such a theory and
how the usual hard objections to trajectories can be avoided.

The conformal parameters and their relation to source terms are an
important development.  The conformal transformations take a place with the
coordinate transformations as  a means of generating a principle of
equivalence.  The usual curvilinear transformations allow the interchange
of gravitational and inertial forces.  The quantum and electrodynamic
forces are now to be included.  The conformal factors are essential for the
description in
the frame of the neutral observer.  These are also important
in the study of the field equations that begins in a
following article.

\section{SUMMARY}
\label{sum}

A number of concepts have been developed which allow the description
of quantum phenomena to be done with differential geometry.
The application is to the combined effects of gravity, electromagnetism
and quantum mechanics without weak or strong interactions.

A number of currently accepted beliefs are discounted, particularly
those concerning the epistemology of
quantum mechanics and the metaphysical basis of general relativity.
A new way of thinking about these fields of physics is devised.
The general theory of relativity must be treated as a phenomenological
result of a deeper theory.  The quantum mechanical theory must not
be derived from classical physics.  A fundamental geometrical
electrodynamics is introduced.

The actual fifth dimension is associated with the proper time of
charged, isolated, point particles of finite mass.  Dependence
on the fifth parameter is not apparent to a real observer because the
physical laws, when expressed in five dimensional form, predict
precisely what would be expected according to common observation.
The proper values of the fifth coordinate of a particle are not absolutely
determinate but are only defined differentially and with a gauge factor
that is not observable.
This approach might be called kinematic dimensional reduction.

The classical principle of equivalence is extended.  This
concept, required by the fundamental notion of trajectory motion, provides
a guide for mathematical development.
An extension of the geometry either as to the
number of dimensions or the use of non-Riemannian effects is required.  The
distinction between the effect of the three fields on a primitive particle
is only defined after the relation between the particle and a neutral
space-time observer is specified.

Fixed gauge methods are used. The individual geometrical fields,
are not subject to the variability of most common
gauge transformations.  The vector potential represents both
the velocity of the
particle and the relationship of the five space to space-time.
Null geodesics are everywhere tangent to the associated
probability current and can represent a quantum state.
The usual problems with Von Neumann's theorem and Bell's theorem
are avoided.  The geometrical theories are in agreement with experimental
results that have been found to limit other alternate quantum theories.
The implications for a five dimensional geometrical theories are
profound and extensive.

\vfill \eject
\begin{figure}
\widetext
\caption{A charged particle and a
neutral particle start in coincidence at $P_1$.  They move
through an interaction
region where they separate onto distinct trajectories.  If
these trajectories be allowed to reintersect at $P_2$,
the principle of equivalence for both particles cannot be valid if attempted
in space-time.   A geometrical extension is needed. \label{f1} }
\end {figure}
\begin{figure}
\widetext
\caption{A quantum clock provides a universal measure of time in a
quantum-gravitational-electromagnetic theory.  A fixed but identified
real particle is transmitted with a nominal fixed speed
through a diffraction apparatus.  By adjusting the slit size, the
character of the diffraction pattern can be observed on the screen.
For an arbitrary but fixed angular
size of the pattern, the slit width provides a dimension
usable for construction of a light pulse clock. \label{f2} }
\end {figure}
\begin{figure}
\narrowtext
\caption{The usual determination of the spacelike distance between two points
fails in a fundamentally quantum theory.  The separation $x_A-x_B$ is to
be found by measuring the transit delays of light beams between the two
particles that move through the points $A$ and $B$.  The photon timing
provides a well defined quantity but the
straightness of the intermediate trajectory of particle A is not established
from conventional quantum phenomenology.\label{f3}  }
\end {figure}
\begin{figure}
\widetext
\caption{A particle is projected through a series of unaligned apertures.
If the apertures are sufficiently small but finite, some of the particles
will traverse the experiment and strike the screen.  The boundary conditions
are critical.  If one or more particles get through, then they must be
supposed to move on trajectories that cannot be described by classical
mechanics. \label{f4} }
\end {figure}
\end{document}